\documentclass[preprintnumbers,noshowpacs,10pt,twocolumn]{revtex4}%
\usepackage{amssymb}
\usepackage{amsfonts}
\usepackage{amsmath}
\usepackage{graphicx}
\usepackage{dcolumn}
\usepackage{bm}
\usepackage{revsymb}%
\begin{document}
\title{Edge of chaos as critical local symmetry breaking in dissipative nonautonomous systems}
\author{Ricardo Chac\'{o}n}
\affiliation{Departamento de F\'{\i}sica Aplicada, E.I.I., Universidad de Extremadura,
Apartado Postal 382, 06006 Badajoz, Spain and Instituto de Computaci\'{o}n
Cient\'{\i}fica Avanzada (ICCAEx), Universidad de Extremadura, 06006 Badajoz,
Spain }
\date{\today}

\begin{abstract}
The fully nonlinear notion of resonance$-$\textit{geometrical resonance}$-$in
the general context of dissipative systems subjected to spatially periodic
\textit{phase-modulated} potentials is discussed. It is demonstrated that
there is an exact local invariant associated with each geometrical resonance
solution which reduces to the system's energy when the potential is
stationary. The geometrical resonance solutions represent a \textit{local
symmetry} whose critical breaking leads to a new analytical criterion for the
onset of chaotic instabilities. This physical criterion is deduced in the
co-moving frame from the local energy conservation over the shortest
significant timescale. Remarkably, the new physical criterion for the onset of
chaotic instabilities is shown to be valid over large regions of parameter
space, thus being useful beyond the scope of current mathematical techniques.
More importantly, the present theory helps to understand the unreasonable
effectiveness of the Melnikov's method beyond the perturbative regime.

\end{abstract}
\maketitle

\section{\bigskip INTRODUCTION}

Hamiltonian and dissipative systems have traditionally been studied separately
due to their clearly different dynamic properties [1]: dissipation forces give
rise to the existence of transient dynamics associated with the basins of the
different attractors, while the Poincar\'{e} integral invariants of
Hamiltonian systems lead to special behaviour of the eigenvalues of equilibria
and periodic orbits, and to existence theorems for various types of orbits
such as the celebrated Kolmogorov-Arnold-Moser theorem. To date, only the
notion of geometrical resonance (GR) [2] has been able to provide a deep link
between autonomous Hamiltonian and non-autonomous dissipative systems in the
sense that it offers a universal procedure with which to locally
"Hamiltonianize" an otherwise dissipative system by suitably choosing the
non-autonomous term(s) $f_{i}(t)$ such that the system's energy is conserved
locally: $f_{i}(t)=f_{i,GR}(t)$. The original formulation of GR analysis was
for \textit{standing }potentials [2], and was applied to diverse nonlinear
problems involving such potentials [3-11], including the stability of the
responses of an overdamped bistable system [7], the suppression of
spatio-temporal chaos and the stabilization of localized solutions in general
spatio-temporal systems [5,8-10], quantum control in trapped Bose-Einstein
condensates (BECs) [11], and the characterization of periodic solutions of a
fractional Duffing's equation [12].

On the other hand, a major body of research has considered \textit{modulated}
potentials appearing in different physical contexts such as synchrotron motion
of beams [13], BECs in optical lattices [14], Shapiro steps and chaos in
Frenkel-Kontorova chains [15], and nanoscale devices powered by the lateral
Casimir force [16], just to cite a few representative examples. In general,
the reference frames co-moving with modulated potentials are accelerating
frames, which introduces an additional complexity into analysis of the
dynamics relative to the laboratory reference frame (L-frame). Since GR is
neither more nor less than a \textit{local symmetry}, namely that the dynamics
equations remain locally invariant under time reversal when the non-autonomous
terms are suitably locally chosen, $f_{i}(t)=f_{i,GR}(t)$, it seems
appropriate and pertinent to explore its implications in general systems.

The article is structured as follows. In Sec. II, new properties of this
subtle symmetry in the generalized context of dissipative systems in
phase-modulated potentials are characterized and exploited to determine a
physical criterion for the onset of chaotic instabilities in parameter space
whose accuracy and scope goes beyond current perturbative mathematical
techniques. The theory is discussed through the paradigmatic model of a
particle subjected to a spatially periodic and temporally shaken potential
which describes, for example, the chaotic phase oscillation of a proton beam
in a cooler synchrotron [13]. Section III includes a summary of the main
results and conclusions. Some details of the analytical calculations are
relegated to the Appendices.

\section{GEOMETRICAL RESONANCE ANALYSIS}

\subsection{Theory}

Let us consider the class of non-autonomous dissipative systems%
\begin{equation}
\overset{..}{x}=g\left(  x,t\right)  -d\left(  x,\overset{.}{x}\right)  ,
\tag{1}%
\end{equation}
where the overdot denotes $d/dt$, $g\left(  x,t\right)  \equiv-\partial
V\left(  x,t\right)  /\partial x$, with $V\left(  x,t\right)  \equiv V\left[
x-f(t)\right]  $ being a modulated and spatially periodic potential while
$f\left(  t\right)  $ is an \textit{a priori} arbitrary (twice-differentiable)
function of time, and where $-d\left(  x,\overset{.}{x}\right)  $ is a generic
dissipative force. In the potential reference frame (V-frame) with $z\left(
t\right)  =x\left(  t\right)  -f(t)$, Eq.~(1) reads%
\begin{equation}
\overset{..}{z}=g\left(  z\right)  -d\left(  z,\overset{.}{z},f,\overset{.}%
{f},t\right)  -\overset{..}{f}. \tag{2}%
\end{equation}
In general, if $z_{GR}(t)$ is a GR solution of Eq.~(2), it must satisfy%
\begin{align}
\overset{..}{z}_{GR}  &  =g\left(  z_{GR}\right)  ,\tag{3}\\
\overset{..}{f}_{GR}  &  =-d\left(  z_{GR},\overset{.}{z}_{GR},f_{GR}%
,\overset{.}{f}_{GR},t\right)  , \tag{4}%
\end{align}
and hence
\begin{equation}
x_{GR}\equiv z_{GR}+f_{GR} \tag{5}%
\end{equation}
is a GR solution of Eq.~(1) for a given set of initial conditions $\left[
x_{0}\equiv x(t=0),\overset{.}{x}_{0}\equiv\overset{.}{x}\left(  t=0\right)
\right]  $. After assuming that $V(z)$ is infinitely differentiable,
definitions (3)-(5) give rise to the following distinguishing properties.
First, in contrast with the case of standing potentials [2], where $f_{GR}$ is
univocally determined from an algebraic equation involving the single GR
solution associated with a given set of initial conditions, one has to solve a
differential equation for $f_{GR}$ in the present general case and obtain the
initial values $f_{GR}\left(  t=0\right)  \equiv f_{GR,0},\overset{.}{f}%
_{GR}(t=0)\equiv\overset{.}{f}_{GR,0}$ as a part of the whole solution. This
is because the GR scenario for modulated potentials involves \textit{two}
reference frames, the V-frame being non-inertial in the general case. Second,
conditions (3)-(4) are equivalent to the local (i.e., dependent on the initial
conditions) energy conservation requirement
\begin{equation}
(1/2)\overset{.}{z}_{GR}^{2}+V\left[  z_{GR}\right]  \equiv E_{GR}=const
\tag{6}%
\end{equation}
in the V-frame, while one has the requirement of a different local invariant
in the L-frame:%
\begin{equation}
I_{GR}\equiv\frac{\overset{.}{x}_{GR}^{2}}{2}+V\left[  x_{GR}-f_{GR}\right]
+\overset{.}{f}_{GR}\left(  \frac{\overset{.}{f}_{GR}}{2}-\overset{.}{x}%
_{GR}\right)  =const. \tag{7}%
\end{equation}
After Taylor expanding the potential, the local invariant $I_{GR}$ can be
recast into the more transparent form%
\begin{equation}
(1/2)\overset{.}{x}_{GR}^{2}+V\left[  x_{GR}-f_{GR,0}\right]  +S_{GR}=const,
\tag{8}%
\end{equation}
where,
\begin{align}
S_{GR}  &  \equiv\sum_{n=1}^{\infty}\frac{\left(  -1\right)  ^{n}}{n!}%
(f_{GR}^{\ }-f_{GR,0})^{n}V^{\left(  n\right)  }\left[  x_{GR}-f_{GR,0}\right]
\nonumber\\
&  +\overset{.}{f}_{GR}\left(  \overset{.}{f}_{GR}/2-\overset{.}{x}%
_{GR}\right)  , \tag{9}%
\end{align}
with $V^{\left(  n\right)  }\equiv d^{n}V(z)/dz^{n}$. From Eq.~(8) one sees
that, under GR conditions, the energy associated with the corresponding
standing potential is \textit{not} (locally) conserved in the L-frame, as
expected, while the new invariant $I_{GR}$ allows the temporal evolution of
this energy to be calculated for each GR solution. Third, in the Hamiltonian
limiting case, i.e., $d\left(  x,\overset{.}{x}\right)  \rightarrow0$, Eq. (4)
becomes $\overset{..}{f}_{GR}=0$, and hence $f_{GR}(t)=Vt$ with $V$ being an
arbitrary constant and where an additional additive constant has been taken to
be zero without loss of generality. This means that, in the absence of
dissipation, GR solutions are solely possible for potentials traveling with
constant speed, i.e., for inertial frames, as expected. And fourth, a GR
solution will be observed if it is stable, i.e., if any small perturbation
$\delta z\left(  t\right)  $ of $z_{GR}\left(  t\right)  $ is damped. After
substituting $z(t)=z_{GR}\left(  t\right)  +\delta z\left(  t\right)  $ into
Eq.~(2) with $f\left(  t\right)  =f_{GR}\left(  t\right)  $, one obtains the
linearized equation of motion for small perturbations $\delta z\left(
t\right)  $:%
\begin{equation}
\overset{..}{\delta z}+\left(  \frac{\partial d}{\partial\overset{.}{z}%
}\right)  _{GR}\overset{.}{\delta z}+\left[  \left(  \frac{\partial
d}{\partial z}\right)  _{GR}-\left(  \frac{dg}{dz}\right)  _{GR}\right]
\delta z=0. \tag{10}%
\end{equation}
Note that this generalized Hill equation with dissipation also governs the
stability of the GR solutions in the L-frame (cf.~Eq.~(5)). It is shown below
that this stability analysis together with the dependence of the GR solutions
on the system's parameters and the local invariants (6) and (8) allows one to
get a new analytical criterion for the order-chaos threshold from the weakest
useful approximation to the local energy conservation in the V-frame.

\subsection{Paradigmatic model}

To demonstrate the effectiveness of the present GR theory in a simple
paradigmatic model, consider the dissipative dynamics of a particle subjected
to a spatially periodic and temporally shaken potential:%
\begin{equation}
\overset{..}{x}+\sin\left[  x-f(t)\right]  =-\eta\overset{.}{x}. \tag{11}%
\end{equation}
The dimensionless Eq.~(11) describes for example the pinion motion of a
nanoscale device composed of a pinion and a rack coupled via the lateral
Casimir force, where $\eta>0$ is a damping coefficient while $f(t)$ accounts
for the \textit{a priori} arbitrary horizontal motion of the rack [16]. In the
V-frame with $z\left(  t\right)  =x\left(  t\right)  -f(t)$, Eq.~(11) reads%
\begin{equation}
\overset{..}{z}+\sin z=-\eta\overset{.}{z}-\eta\overset{.}{f}-\overset{..}{f}.
\tag{12}%
\end{equation}
Thus, GR solutions of Eq.~(12) must satisfy%
\begin{align}
\overset{..}{z}_{GR}+\sin z_{GR}  &  =0,\tag{13}\\
\overset{.}{z}_{GR}+\overset{.}{f}_{GR}+\overset{..}{f}_{GR}/\eta &  =0.
\tag{14}%
\end{align}
Exact analytical periodic solutions of the integrable pendulum (13) [17]
corresponding to libration and rotation motions are given by
\begin{align}
z_{GR}^{l}(t;t_{0},m)  &  =2\arcsin\left[  \sqrt{m}\operatorname*{sn}\left(
t-t_{0};m\right)  \right]  ,\nonumber\\
\overset{.}{z}_{GR}^{l}(t;t_{0},m)  &  =2\sqrt{m}\operatorname{cn}\left(
t-t_{0};m\right)  , \tag{15}%
\end{align}
and%
\begin{align}
z_{GR}^{r}(t;t_{0},m)  &  =\pm2\operatorname{am}\left[  \left(  t-t_{0}%
\right)  /\sqrt{m};m\right]  ,\nonumber\\
\overset{.}{z}_{GR}^{r}(t;t_{0},m)  &  =\pm\frac{2}{\sqrt{m}}%
\operatorname*{dn}\left[  \frac{t-t_{0}}{\sqrt{m}};m\right]  , \tag{16}%
\end{align}
respectively, where $\operatorname*{sn}\left(  \cdot;m\right)  $,
$\operatorname{cn}\left(  \cdot;m\right)  $, $\operatorname*{dn}\left(
\cdot;m\right)  $, $\operatorname{am}\left(  \cdot;m\right)  $ are Jacobian
elliptic functions of parameter $m\in\left]  0,1\right[  $, the upper (lower)
sign in the rotation solutions refers to counterclockwise (clockwise)
rotations, while $t_{0}$ is an arbitrary initial time. These solutions have
the respective periods $T^{l}\left(  m\right)  \equiv4K$ and $T^{r}%
(m)\equiv2\sqrt{m}K$, where $K\equiv K(m)$ is the complete elliptic integral
of the first kind [18]. Although the parameters corresponding to libration and
rotation motions are inversely related each other, the same notation, $m$, is
used here since both parameters are defined over the same interval [17] and
there is no possibility of confusion in the subsequent analysis. Also,
definition $\operatorname*{sn}\left(  \cdot;m\right)  \equiv\sin\left[
\operatorname{am}\left(  \cdot;m\right)  \right]  $ has been used to write a
simpler alternative expression for $z_{GR}^{r}(t;t_{0},m)$. After taking
$t_{0}=0$ for simplicity, using the Fourier series of the Jacobian elliptic
functions involved [18], and integrating the linear differential equation
(14), one straightforwardly obtains the GR excitations%
\begin{align}
f_{GR}^{l}(t)  &  \equiv C_{1}+C_{2}\operatorname{e}^{-\eta t}+\sum
_{n=0}^{\infty}b_{n}\cos\left(  \omega_{n}t+\varphi_{n}\right)  ,\nonumber\\
f_{GR}^{r}(t)  &  \equiv C_{1}^{\prime}+C_{2}^{\prime}\operatorname{e}^{-\eta
t}\mp\frac{\pi t}{\sqrt{m}K}\pm\sum_{n=1}^{\infty}b_{n}^{\prime}\cos\left(
\omega_{n}^{\prime}t+\varphi_{n}^{\prime}\right)  , \tag{17}%
\end{align}
and the corresponding GR solutions in the L-frame
\begin{align}
x_{GR}^{l}(t)  &  =C_{1}+C_{2}\operatorname{e}^{-\eta t}+\sum_{n=0}^{\infty
}\frac{b_{n}}{\operatorname*{tg}\varphi_{n}}\sin\left(  \omega_{n}%
t+\varphi_{n}\right)  ,\nonumber\\
x_{GR}^{r}(t)  &  =C_{1}^{\prime}+C_{2}^{\prime}\operatorname{e}^{-\eta t}%
\pm\sum_{n=1}^{\infty}\frac{b_{n}^{\prime}}{\operatorname*{tg}\varphi
_{n}^{\prime}}\sin\left(  \omega_{n}^{\prime}t+\varphi_{n}^{\prime}\right)  ,
\tag{18}%
\end{align}
with
\begin{align}
b_{n}  &  \equiv\frac{2\pi\eta}{\omega_{n}K\sqrt{\eta^{2}+\omega_{n}^{2}}%
}\operatorname{sech}\left[  \frac{\left(  2n+1\right)  \pi K^{\prime}}%
{2K}\right]  ,\nonumber\\
b_{n}^{\prime}  &  \equiv\frac{2\pi\eta}{\omega_{n}^{\prime}K\sqrt{m}%
\sqrt{\eta^{2}+\omega_{n}^{\prime2}}}\operatorname{sech}\left(  \frac{n\pi
K^{\prime}}{K}\right)  ,\nonumber\\
\omega_{n}  &  \equiv\left(  n+1/2\right)  \pi/K,\nonumber\\
\omega_{n}^{\prime}  &  \equiv n\pi/\left(  \sqrt{m}K\right)  ,\nonumber\\
\varphi_{n}  &  \equiv\arctan\left(  \eta/\omega_{n}\right)  ,\nonumber\\
\varphi_{n}^{\prime}  &  \equiv\arctan\left(  \eta/\omega_{n}^{\prime}\right)
\tag{19}%
\end{align}
and where $C_{1,2},C_{1,2}^{\prime}$ are constants to be determined from the
initial conditions $\left(  x_{0},\overset{.}{x}_{0}\right)  $ (see Appendix
A), $K^{\prime}\equiv K(1-m)$, while the upper (lower) sign in Eqs.~(17) and
(18) refers to counterclockwise (clockwise) rotations. These GR solutions have
the following properties. (i) Their stability is governed by Eq.~(10), i.e.,
\begin{equation}
\overset{..}{\delta z}+\eta\overset{.}{\delta z}+\cos z_{GR}^{l,r}\delta z=0,
\tag{20}%
\end{equation}
which reduces to the Lam\'{e} equations%
\begin{align}
\frac{d^{2}u}{dt^{2}}+\left[  1-\eta^{2}/4-2m\operatorname{sn}^{2}\left(
t;m\right)  \right]  u  &  =0,\tag{21}\\
\frac{d^{2}v}{d\tau^{2}}+\left[  m\left(  1-\eta^{2}/4\right)
-2m\operatorname{sn}^{2}\left(  \tau;m\right)  \right]  v  &  =0, \tag{22}%
\end{align}
where $u=\exp\left(  \eta t/2\right)  \delta z$ and $v=\exp\left(  \eta
\sqrt{m}\tau/2\right)  \delta z$ for librations and rotations, respectively.
Standard results for these Lam\'{e} equations [19,20] indicate that Eq.~(20)
presents only one instability region in the $m-\eta$ parameter plane. A
careful comparison of Eq.~(21) with Eq.~(22) leads one to expect the
instability region for librations to be clearly narrower than that for
rotations owing to the term $m\left(  1-\eta^{2}/4\right)  <1-\eta^{2}/4$
since $m\in\left]  0,1\right[  $. Moreover, the maximum range of $\eta$ values
in the instability regions is expected to occur when $m\simeq1$ for both kinds
of motion due to all GR solutions $z_{GR}^{l,r}\left(  t\right)  $ converging
to the separatrix (the most unstable phase path) of the integrable pendulum as
$m\rightarrow1$. Numerical simulations confirmed these expectations, as is
shown in Fig.~1. (ii) For any set of initial conditions not on the unperturbed
separatrix, i.e., for \textit{any} GR excitation (17) and corresponding
solution (18), one sees that the dependence of \textit{each} harmonic of such
excitations and solutions on the damping coefficient has exactly the same
form: $\eta/\sqrt{\eta^{2}+\alpha}$, with $\alpha$ being a function of the
corresponding natural period. From this it can be inferred that, for a
periodic excitation $f(t)$ of amplitude $\gamma$, the dependence of the
chaotic-threshold amplitude, $\gamma_{th}$, on $\eta$ should obey this
functional form \textit{irrespective }of the value of $\eta$, an unanticipated
result in view of the \textit{perturbative} character of the current
mathematical techniques to predict the onset of chaos (Melnikov's method (MM)
[21]). (iii) Since harmonic functions are commonly used to model periodic
excitations, the GR solution corresponding to libration near the bottom of the
potential well (i.e., $x_{0},\approx0,\overset{.}{x}_{0}\approx0,m\gtrsim0$)
is of especial interest. One straightforwardly obtains the steady $(t\gg
\eta^{-1})$ solutions (cf. Eqs.~(17) and (18))%
\begin{align}
x_{GR}^{l}(t)  &  =2\sqrt{m}\sin t+f_{GR}^{l}(t),\nonumber\\
f_{GR}^{l}(t)  &  =\frac{2\sqrt{m}\eta}{\sqrt{1+\eta^{2}}}\cos\left(
t+\arctan\eta\right)  +O\left(  m^{3/2}\right)  , \tag{23}%
\end{align}
while the corresponding local invariant (8) reduces to%
\begin{align}
I_{GR}^{l}  &  =(1/2)\overset{.}{x}_{GR}^{l2}\left(  t\right)  -\cos
x_{GR}^{l}\left(  t\right)  +\frac{f_{GR}^{l2}(t)+\overset{.}{f}_{GR}^{l2}%
(t)}{2}\nonumber\\
&  +O\left(  m^{3}\right)  , \tag{24}%
\end{align}
i.e., $I_{GR}^{l}\left(  t\right)  $ is no more than the sum of the energy
associated with the limiting case of standing potential plus the energy
associated with the V-frame moving \textit{as a linear harmonic oscillator of
period }$2\pi$, which is an unexpected result. (iv) In the Hamiltonian
limiting case, i.e., $\eta\rightarrow0$, GR solutions for librations are not
possible due to their oscillatory character around a fixed point is
incompatible with the requirement of a traveling potential function (cf. third
property in the previous subsection), while GR solutions for rotations are
indeed possible (see Eq. (17)). (v) In the limit of very high dissipation
$\left(  \eta\rightarrow\infty\right)  $, the steady GR solutions are
equilibria (cf.~Eqs.~(18) and (23)), as expected.

\begin{figure}[tbh]
\includegraphics[width=0.3\textwidth]{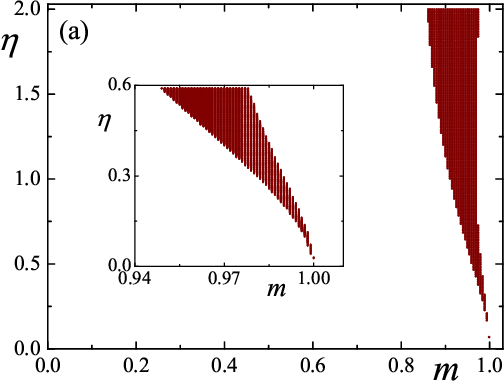}
\includegraphics[width=0.3\textwidth]{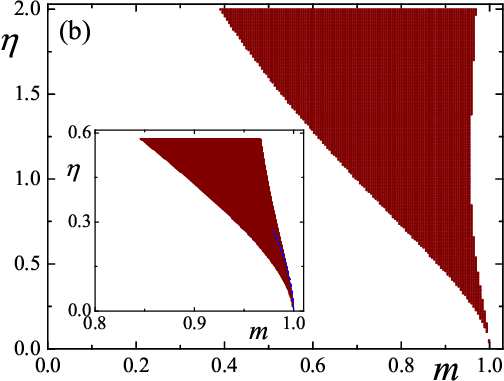}
\caption{Stability-instability charts obtained by numerical
integration of the
Lam\'{e}'s equations (a) Eq.~(21) and (b) Eq.~(22) for libration and
rotation
motions, respectively, where instability regions are indicated by
dots. The
insets show the tips of the instability tongues when $m\rightarrow1$ and
$\eta\rightarrow0$.}
\label{fig1}
\end{figure}

\subsection{Order-chaos threshold}

Next, one can use the above properties of the GR solutions to obtain an
analytical estimate of the order-chaos threshold associated with a generic
$T$-periodic excitation $f_{g}(t)$ of amplitude $\gamma$. In general this
generic excitation will not exactly correspond to any GR excitation function
(17), and hence one cannot expect strict conservation (i.e., over an infinite
timescale) of the invariants (6) and (7) for any set of initial conditions.
Indeed, the energy rate is governed in the V-frame by the equation (cf.~Eq.
(12))%
\begin{equation}
\frac{dE}{dt}=-\eta\overset{.}{z}^{2}-\eta\overset{.}{z}\overset{.}{f_{g}%
}-\overset{.}{z}\overset{..}{f}_{g}. \tag{25}%
\end{equation}
For each set of initial conditions, the closer the excitation $f_{g}(t)$ is to
the corresponding $f_{GR}(t)$, the smaller the deviation of the energy from
the corresponding local invariant $E_{GR}$. Clearly, the weakest physical
condition that will cope with this deviation is that the energy be locally
conserved over the shortest significant timescale, i.e., as an average over a
period of the corresponding GR solution:%
\begin{align}
\left\langle \frac{dE}{dt}\right\rangle _{T^{l,r}}\left(  t_{0}\right)   &
\equiv\int_{-T^{l,r}/2}^{T^{l,r}/2}\left(  \eta\overset{.}{z}_{GR}^{l,r2}%
+\eta\overset{.}{z}_{GR}^{l,r}\overset{.}{f_{g}}+\overset{.}{z}_{GR}%
^{l,r}\overset{..}{f}_{g}\right)  dt\nonumber\\
&  =0 \tag{26}%
\end{align}
for some $t_{0}$, and where $\overset{.}{z}_{GR}^{l,r}\equiv\overset{.}%
{z}_{GR}^{l,r}\left(  t;t_{0},m\right)  $. Also, one assumes a
\textit{Galilean} resonance condition$-$a necessary condition for GR
(cf.~Eq.~(14))$-$for both libration ($T=T^{l}/(2n+1)$ for some $n=0,1,...$)
and rotation ($T=T^{r}/n$ for some $n=1,2,...$) motions. Thus, Eq.~(26)
provides a \textit{local} condition that takes into account the initial phase
difference between the generic excitation and the GR solution, hence allowing
one to obtain a threshold condition (in particular, a threshold amplitude
$\gamma_{th}$) for the energy conservation in its weakest sense. According to
the above stability analysis, GR solutions are not uniformly stable as the
natural period is varied. Therefore, Eq.~(26) is subject to the caveat that it
is not expected to be uniformly valid for all values of the excitation period
because of its dependence on the integration domain. In the limiting case
$T^{l,r}\rightarrow\infty$, when both libration and rotation GR solutions
converge to the separatrix
\begin{align}
z_{s,\pm}(t;t_{0})  &  =\pm\arctan\left[  \sinh\left(  t-t_{0}\right)
\right]  ,\nonumber\\
\overset{.}{z}_{s,\pm}\left(  t;t_{0}\right)   &  =\pm2\operatorname{sech}%
\left(  t-t_{0}\right)  , \tag{27}%
\end{align}
the corresponding GR excitation is no longer a periodic function, as expected,
but (for $t_{0}=0$)
\begin{align}
f_{GR}^{s}(t)  &  \equiv\widetilde{C}_{1}+\widetilde{C}_{2}\operatorname{e}%
^{-\eta t}\pm4\operatorname{e}^{t}\frac{_{2}F_{1}(1,\frac{1+\eta}{2}%
;\frac{3+\eta}{2};-\operatorname{e}^{2t})}{1+\eta}\nonumber\\
&  \mp2\operatorname*{gd}\left(  t\right)  , \tag{28}%
\end{align}
where $\widetilde{C}_{1,2}$ are constants to be determined from the initial
conditions $\left(  x_{0},\overset{.}{x}_{0}\right)  $ (see Appendix B),
$\operatorname*{gd}\left(  t\right)  $ and $_{2}F_{1}(1,\frac{1+\eta}{2}%
;\frac{3+\eta}{2};-\operatorname{e}^{2t})$ are the Gudermannian and the
hypergeometric functions, respectively [22], and Eq.~(26) becomes%
\begin{equation}
\left\langle \frac{dE}{dt}\right\rangle _{s}\left(  t_{0}\right)  \equiv
\int_{-\infty}^{\infty}\left(  \eta\overset{.}{z}_{s,\pm}^{2}+\eta\overset
{.}{z}_{s,\pm}\overset{.}{f_{g}}+\overset{.}{z}_{s,\pm}\overset{..}{f}%
_{g}\right)  dt=0 \tag{29}%
\end{equation}
for some $t_{0}$. Since the separatrix is the most unstable phase path, in the
sense that it is the boundary between two \textit{distinctly different} types
of motions, one would expect the onset of chaotic instabilities when a gradual
breaking of the GR local symmetry reaches a critical value. Indeed, Eq.~(29)
provides the physical condition for such a critical breaking, hence allowing
the order-chaos threshold in parameter space to be estimated. It should be
stressed that, because the GR local symmetry is defined over the
\textit{complete} parameter space, condition (29) is postulated irrespective
of the parameter values. For the sake of clarity, consider the application of
condition (26) to the simple case of a harmonic excitation $f(t)=\gamma
\cos\left(  2\pi t/T\right)  $. After some simple algebra (see Appendix C),
one straightforwardly obtains the following threshold amplitudes from
Eq.~(26):
\begin{align}
\gamma_{th}^{l}  &  =\frac{4\eta\left[  E^{\ast}-(1-m)K\right]  }{\pi
\omega\sqrt{\eta^{2}+\omega^{2}}\operatorname{sech}\left(  \omega K^{\prime
}\right)  },\tag{30}\\
\gamma_{th}^{r}  &  =\frac{4\eta\sqrt{m}E^{\ast}}{\pi\omega\sqrt{\eta
^{2}+\omega^{2}}\operatorname{sech}\left(  \omega\sqrt{m}K^{\prime}\right)  },
\tag{31}%
\end{align}
where $\omega\equiv2\pi/T$ and $E^{\ast}\equiv E^{\ast}(m)$ is the complete
elliptic integral of the second kind [18]. Also, $\lim_{m\rightarrow1}%
\gamma_{th}^{l,r}=\gamma_{th}^{s}$ with%
\begin{equation}
\gamma_{th}^{s}\equiv\frac{4\eta}{\pi\omega\sqrt{\eta^{2}+\omega^{2}%
}\operatorname{sech}\left(  \pi\omega/2\right)  } \tag{32}%
\end{equation}
being the explicit estimate of the order-chaos threshold in parameter space,
while
\begin{equation}
\gamma\geqslant\gamma_{th}^{s} \tag{33}%
\end{equation}
provides a \textit{necessary} condition for the onset of chaotic (homoclinic)
instabilities (cf. Appendix C) which comes from a \textit{physical} condition:
the critical breaking of the local conservation of the separatrix's energy in
the co-moving frame.

\subsection{Comparison with Melnikov's method results}

Let us now compare the prediction (32) with that obtained from MM. In keeping
with the assumptions of the MM [23,1,21], here it is assumed that one can
write $\eta=\varepsilon\overline{\eta},\gamma=\varepsilon\overline{\gamma}$
where $0<\varepsilon\ll1$ and $\overline{\eta},\overline{\gamma},\omega$ are
of order unity. Next, one calculates the Melnikov function (MF), $M(t_{0})$,
for the system (12) with the harmonic excitation:%
\begin{equation}
\overset{..}{z}+\sin z=-\varepsilon\overline{\eta}\overset{.}{z}%
+\varepsilon\overline{\gamma}\omega^{2}\cos\left(  \omega t\right)  +O\left(
\varepsilon^{2}\right)  ,\tag{34}%
\end{equation}
with $O\left(  \varepsilon^{2}\right)  \equiv\varepsilon^{2}\overline{\eta
}\overline{\gamma}\omega\sin\left(  \omega t\right)  $. Since the MF provides
an $O\left(  \varepsilon\right)  $ estimate of the distance between the stable
and unstable manifolds of the perturbed system in the Poincar\'{e} section at
$t_{0}$, one readily obtains
\begin{equation}
M(t_{0})=-8\eta\pm\frac{2\pi\gamma\omega^{2}}{\cosh\left(  \pi\omega/2\right)
}\cos\left(  \omega t_{0}\right)  +O\left(  \varepsilon^{2}\right)  .\tag{35}%
\end{equation}
If the MF has a simple zero, then a homoclinic bifurcation occurs, signifying
the appearance of chaotic (homoclinic) instabilities [24]. This yields the
threshold value%
\begin{equation}
\gamma_{th,\varepsilon}^{M}\equiv\frac{4\eta}{\pi\omega^{2}\operatorname{sech}%
\left(  \pi\omega/2\right)  },\tag{36}%
\end{equation}
while
\begin{equation}
\gamma\geqslant\gamma_{th,\varepsilon}^{M}\tag{37}%
\end{equation}
provides a \textit{necessary} condition for the onset of chaotic (homoclinic)
instabilities which comes from a \textit{mathematical} condition: the
intersection between the stable and unstable manifolds of the perturbed system
in the Poincar\'{e} section at some $t_{0}$, which is deduced from the
analysis of an $O\left(  \varepsilon\right)  $ estimate of the distance
between such manifolds according to MM. Surprisingly, if one forces retaining
the term $\varepsilon^{2}\overline{\eta}\overline{\gamma}\omega\sin\left(
\omega t\right)  $ (cf. Eq. (34)) when calculating the MF, one readily obtains%
\begin{equation}
M(t_{0})=-8\eta\pm\frac{2\pi\gamma\omega\left[  \omega\cos\left(  \omega
t_{0}\right)  +\eta\sin\left(  \omega t_{0}\right)  \right]  }{\cosh\left(
\pi\omega/2\right)  }.\tag{38}%
\end{equation}
This yields the new threshold value%
\begin{equation}
\gamma_{th,\varepsilon^{2}}^{M}\equiv\frac{4\eta}{\pi\omega\sqrt{\eta
^{2}+\omega^{2}}\operatorname{sech}\left(  \pi\omega/2\right)  }\equiv
\gamma_{th}^{s},\tag{39}%
\end{equation}
i.e., the same threshold amplitude than that obtained from the critical
breaking of the local conservation of the separatrix's energy in the co-moving
frame (cf. Eq. (32)).

Numerical simulations confirmed the effectiveness of estimate (32), (39) as
against (36). Equation (11) has been numerically solved using a fourth
Runge-Kutta method with discrete time step $\delta t=0.001$. Lyapunov
exponents have been computed using a version of the algorithm introduced in
Ref. [25], with integration typically up to $2\times10^{4}$ drive cycles for
each fixed set of parameters. An illustrative example is shown in Fig.~2, in
which one sees how the chaotic regions in the $\eta-\gamma$ parameter plane,
determined by Lyapunov exponent (LE) calculations, are reasonably well bounded
by estimate (32), while the \textit{extrapolation} (recall that $\eta,\gamma$
must be much smaller than unity) of the MM estimate (36) clearly fails. Note
that estimates (32) and (36) coincide for sufficiently small values of $\eta$
(perturbative regime). This is not so surprising since the MF is, up to a
constant, exactly the integral that Poincar\'{e} derived from Hamilton-Jacobi
theory to obtain his celebrated obstruction to integrability [26], while the
GR local symmetry implies a local restoration of integrability of an otherwise
\textit{non-integrable} system. It is worth mentioning that, even in the case
of sufficiently small values of $\eta$ and $\gamma$, one cannot expect too
good a quantitative agreement between the numerical findings and the
theoretical predictions (Eq. (32)) because LE provides information solely
concerning attractors (steady responses), while the critical breaking of the
local conservation of the separatrix's energy is related to the onset of
chaotic instabilities (i.e., it is generally related to transient chaos).
Figure 3 shows illustrative examples of the regularization routes as $\gamma$
and $\eta$ are changed while crossing the order-chaos threshold (recall the
aforementioned caveat). Typically, the system (11) goes from a period-1
attractor to a strange chaotic attractor as the excitation amplitude increases
for a sufficiently small value of the damping coefficient (see Fig. 3(a)). The
overall evolution of the initial periodic state is characterized by the energy
$E=\overset{.}{x}^{2}/2+1-\cos x$ undergoing a period-doubling route as
$\gamma$ is increased. Also, for fixed $\gamma$, the system (11) goes from the
strange chaotic attractor existing at a sufficiently small value of the
damping coefficient to a period-1 attractor as $\eta$ is increased via an
inverse period-doubling route (see Fig. 3(b)). These numerical findings
confirmed the effectiveness of the estimation (32), (39), providing an
additional instance of the unreasonable effectiveness of MM predictions beyond
the perturbative regime (see, e.g., Ref. [29]).
\begin{figure*}[tbh]
\includegraphics[width=0.25\textwidth]{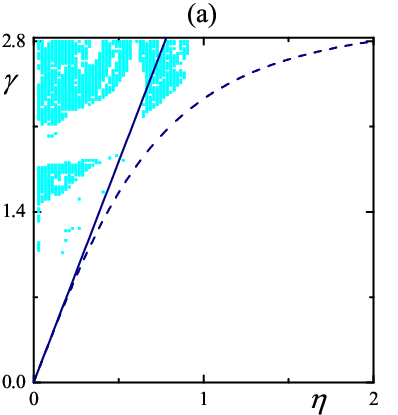}
\includegraphics[width=0.25\textwidth]{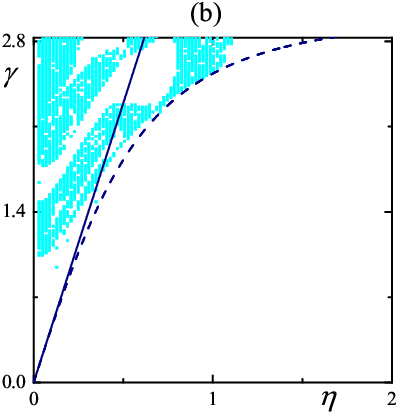}
\includegraphics[width=0.25\textwidth]{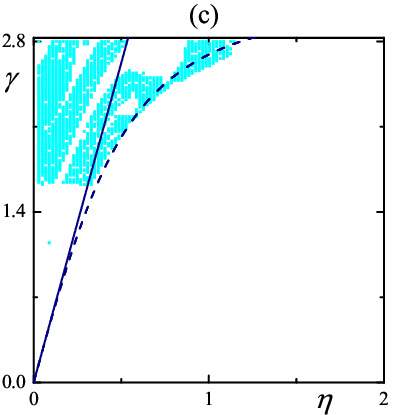}
\caption{\textbf{ }Chaotic regions (dots) in the $\eta-\gamma$
parameter plane
corresponding to Eq.~(11) with $f(t)=\gamma\cos\left(  2\pi t/T\right)
$ for
three values of the driving period: (a) $T=7.5$, (b) $T=9.38$, (c)
$T=10.5$. A
dot is plotted on a $100\times140$ grid when the corresponding maximal
LE is
greater than $10^{-3}$. Dashed and solid lines represent the theoretical
chaotic thresholds [cf.~Eqs.(32) and (36), respectively] from GR
analysis and
MM, respectively.}
\label{fig2}
\end{figure*}

\begin{figure*}[tbh]
\includegraphics[width=0.3\textwidth]{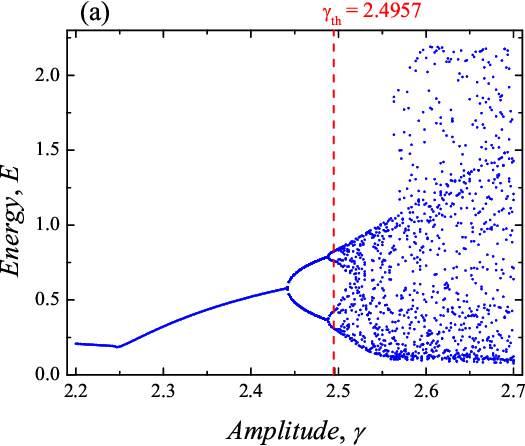}
\includegraphics[width=0.3\textwidth]{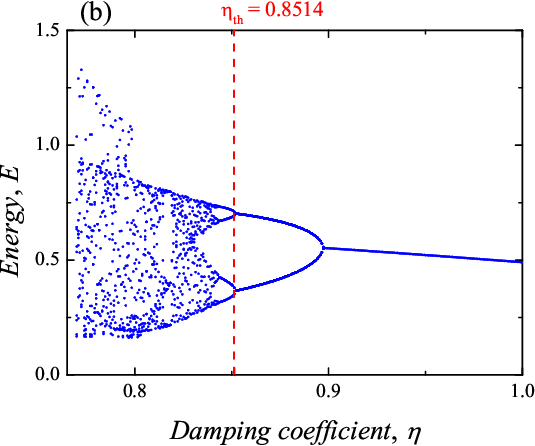}
\caption{Bifurcation diagrams of Eq. (11) with $f(t)=\gamma\cos\left(
2\pi
t/T\right)  $. (a)\textbf{ }Bifurcation diagram of energy
$E=\overset{.}%
{x}^{2}/2+1-\cos x$ as a function of the amplitude $\gamma$ for
$\eta=0.95$.
The vertical dashed line indicates the threshold amplitude
$\gamma_{th}=2.4957$
predicted from Eq. (32). (b) Bifurcation diagram of energy
$E=\overset{.}%
{x}^{2}/2+1-\cos x$ as a function of the damping coefficient $\eta$ for
$\gamma=2.4$. The vertical dashed line indicates the threshold value of
the
damping coefficient $\eta_{th}=0.8514$ predicted from Eq. (32). Fixed
parameter: $T=9.38$.}
\label{fig3}
\end{figure*}

\section{CONCLUSION}

In conclusion, a theory of geometrical resonance in dissipative systems
subjected to spatially periodic phase-modulated potentials has been presented,
and its effectiveness in obtaining an analytical criterion for the onset of
chaotic (homoclinic) instabilities in parameter space beyond the perturbative
regime demonstrated by means of a paradigmatic example. From a theoretical
point of view, the characterization and determination of the frontiers between
chaotic and regular motions of real-world systems is a fundamental problem
that needs to be addressed in all branches of science. While the mathematical
theory of deterministic chaos was definitively established by the work of
Poincar\'{e}, Birkhoff, and Smale, the present physical theory suggests
understanding the onset of homoclinic chaotic instabilities as being
coincident with a \textit{critical} \textit{breaking of the geometrical
resonance local symmetry}, specifically, as coinciding with a critical
breaking of the local conservation of the separatrix's energy in the co-moving
frame for generic spatially periodic phase-modulated potentials. More
importantly, the present theory helps to understand the unreasonable
effectiveness of the predictions of the Melnikov's method beyond the
perturbative regime. Finally, a natural continuation of the present work is
the study of the geometrical resonance local symmetry and its eventual
breakage in classical and quantum Hamiltonian systems.

\section{ACKNOWLEDGEMENT}

The author thanks F. Balibrea, M. Berry, P. Binder, D. Farmer, E.
Hern\'{a}ndez-Garc\'{\i}a, and P. J. Mart\'{\i}nez for discussion and useful
comments on an early version of the manuscript. This work was supported by the
Ministerio de Ciencia, Innovaci\'{o}n y Universidades (MICIU, Spain) through
Project No. PID2019-108508GB-100/AEI/10.13039/501100011033 cofinanced by FEDER
funds and by the Junta de Extremadura (JEx, Spain) through Project No. GR21012
cofinanced by FEDER funds.

\section*{APPENDIX A: DETERMINATION OF THE INITIAL VALUE PROBLEM}

From Eqs. (13), (15), (16) with $t_{0}=0$, one obtains%
\begin{align}
f_{GR,0}^{l,r}  &  \equiv f_{GR}^{l,r}\left(  t=0\right)  =x_{GR}%
^{l,r}(t=0)-z_{GR}^{l,r}(t=0)=x_{0},\nonumber\\
\overset{.}{f}_{GR,0}^{l}  &  \equiv\overset{.}{f}_{GR}^{i}(t=0)=\overset
{.}{x}_{GR}^{l}(t=0)-\overset{.}{z}_{GR}^{l}(t=0)\nonumber\\
&  =\overset{.}{x}_{0}-2\sqrt{m},\nonumber\\
\overset{.}{f}_{GR,0}^{r}  &  \equiv\overset{.}{f}_{GR}^{r}(t=0)=\overset
{.}{x}_{GR}^{r}(t=0)-\overset{.}{z}_{GR}^{r}(t=0)\nonumber\\
&  =\overset{.}{x}_{0}\mp\frac{2}{\sqrt{m}}, \tag{A1}%
\end{align}
while from Eq. (18) one obtains%
\begin{align}
C_{1}  &  =x_{0}+\frac{1}{\eta}\left[  \overset{.}{x}_{0}-\sum_{n=0}^{\infty
}\frac{b_{n}\omega_{n}\cos^{2}\varphi_{n}}{\sin\varphi_{n}}\right]
-\sum_{n=0}^{\infty}b_{n}\cos\varphi_{n},\nonumber\\
C_{2}  &  =\frac{1}{\eta}\left[  -\overset{.}{x}_{0}+\sum_{n=0}^{\infty}%
\frac{b_{n}\omega_{n}\cos^{2}\varphi_{n}}{\sin\varphi_{n}}\right]  , \tag{A2}%
\end{align}
for libration motions, and%
\begin{align}
C_{1}^{\prime}  &  =x_{0}+\frac{1}{\eta}\left[  \overset{.}{x}_{0}\mp
\sum_{n=1}^{\infty}\frac{b_{n}^{\prime}\omega_{n}^{\prime}\cos^{2}\varphi
_{n}^{\prime}}{\sin\varphi_{n}^{\prime}}\right]  \mp\sum_{n=1}^{\infty}%
b_{n}^{\prime}\cos\varphi_{n}^{\prime},\nonumber\\
C_{2}^{\prime}  &  =\frac{1}{\eta}\left[  -\overset{.}{x}_{0}\pm\sum
_{n=1}^{\infty}\frac{b_{n}^{\prime}\omega_{n}^{\prime}\cos^{2}\varphi
_{n}^{\prime}}{\sin\varphi_{n}^{\prime}}\right]  , \tag{A3}%
\end{align}
for rotation motions, and where $b_{n},\omega_{n},\varphi_{n},b_{n}^{\prime
},\omega_{n}^{\prime},\varphi_{n}^{\prime}$ are given by Eq. (19), while the
upper (lower) sign in Eqs.~(A1) and (A3) refers to counterclockwise
(clockwise) rotations.

\section*{APPENDIX B: DERIVATION OF THE GEOMETRICAL RESONANCE EXCITATION FOR
THE SEPARATRIX}

For the separatrix (27), Eq. (14) reduces to the linear differential equation%
\begin{equation}
\overset{..}{f}_{GR}^{s}+\eta\overset{.}{f}_{GR}^{s}=\mp2\eta
\operatorname{sech}\left(  t\right)  . \tag{B1}%
\end{equation}
After using the method of variation of parameters [27], one straightforwardly
obtains the general solution given by Eq. (28) while its derivative is written%
\begin{equation}
\overset{.}{f}_{GR}^{s}(t)=-\widetilde{C}_{2}\eta\operatorname{e}^{-\eta t}%
\mp4\eta\operatorname{e}^{t}\frac{_{2}F_{1}(1,\frac{1+\eta}{2};\frac{3+\eta
}{2};-\operatorname{e}^{2t})}{1+\eta}. \tag{B2}%
\end{equation}
Finally, after taking into account (A1) for $m=1$, i.e., $f_{GR}%
^{s}(t=0)=x_{0},\overset{.}{f}_{GR}^{s}(t=0)=\overset{.}{x}_{0}\mp2$, the
integration constants are given by,%
\begin{align}
\widetilde{C}_{1}  &  =x_{0}+\frac{1}{\eta}\left(  \overset{.}{x}_{0}%
\mp2\right)  ,\nonumber\\
\widetilde{C}_{2}  &  =\frac{1}{\eta}\left\{  -\overset{.}{x}_{0}\pm2\mp
\eta\left[  \psi\left(  \frac{3+\eta}{4}\right)  -\psi\left(  \frac{1+\eta}%
{4}\right)  \right]  \right\}  , \tag{B3}%
\end{align}
where $\psi\left(  \eta\right)  $ is the psi (Digamma) function [22], while
the upper (lower) sign in Eqs.~(28), (B1), (B2), and (B3) refers to
counterclockwise (clockwise) rotations.

\section*{APPENDIX C: DERIVATION OF THE THRESHOLD AMPLITUDES FOR THE ONSET OF
CHAOTIC INSTABILITIES}

Let us consider the simple case of a harmonic excitation $f(t)=\gamma
\cos\left(  2\pi t/T\right)  $.

\textit{Libration motions.} Equation (26) for $T^{l}\left(  m\right)
\equiv4K$ and the Galilean resonance condition $T=T^{l}/(2n+1),n=0,1,...$,
reduces to%
\begin{align}
\left\langle \frac{dE}{dt}\right\rangle _{T^{l}}\left(  t_{0}\right)   &
=\int_{-2K}^{2K}\left[  \eta\overset{.}{z}_{GR}^{l2}-\eta\gamma\omega
\overset{.}{z}_{GR}^{l}\sin\left(  \omega t\right)  \right]  dt\nonumber\\
&  -\int_{-2K}^{2K}\gamma\omega^{2}\overset{.}{z}_{GR}^{l}\cos\left(  \omega
t\right)  dt. \tag{C1}%
\end{align}
After substituting $\overset{.}{z}_{GR}^{l}(t;t_{0},m)$ from Eq. (15) into Eq.
(C1), using the Fourier series of the elliptic function $\operatorname{cn}%
\left(  \cdot;m\right)  $ [18], and using standard tables of integrals [28],
one obtains the average energy over the period $T^{l}$ as a function of the
initial phase difference $\left(  \omega t_{0}\right)  $ between the harmonic
excitation and the GR solution
\begin{align}
\left\langle \frac{dE}{dt}\right\rangle _{T^{l}}\left(  t_{0}\right)   &
=16\eta\left[  E^{\ast}-(1-m)K\right]  -\frac{4\pi\gamma\omega\eta\sin\left(
\omega t_{0}\right)  }{\cosh\left[  \omega K^{\prime}\right]  }\nonumber\\
&  -\frac{4\pi\gamma\omega^{2}\cos\left(  \omega t_{0}\right)  }{\cosh\left[
\omega K^{\prime}\right]  }. \tag{C2}%
\end{align}
From Eq. (C2) one sees that a \textit{necessary} condition for $\left\langle
\frac{dE}{dt}\right\rangle _{T^{l}}\left(  t_{0}\right)  $ to change sign at
some $t_{0}$ is written
\begin{equation}
16\eta\left[  E^{\ast}-(1-m)K\right]  \leqslant\sqrt{\left(  \frac{4\pi
\gamma\omega\eta}{\cosh\left[  \omega K^{\prime}\right]  }\right)
^{2}+\left(  \frac{4\pi\gamma\omega^{2}}{\cosh\left[  \omega K^{\prime
}\right]  }\right)  ^{2}}, \tag{C3}%
\end{equation}
where the equals sign in Eq. (C3) yields the threshold amplitude $\gamma
_{th}^{l}$ given by Eq. (30).

\textit{Rotation motions.} Equation (26) for $T^{r}\left(  m\right)  \equiv$
$2\sqrt{m}K$ and the Galilean resonance condition $T=T^{r}/n$, $n=1,2,...$,
reduces to%
\begin{align}
\left\langle \frac{dE}{dt}\right\rangle _{T^{r}}\left(  t_{0}\right)   &
=\int_{-\sqrt{m}K}^{\sqrt{m}K}\left[  \eta\overset{.}{z}_{GR}^{r2}-\eta
\gamma\omega\overset{.}{z}_{GR}^{r}\sin\left(  \omega t\right)  \right]
dt\nonumber\\
&  -\int_{-\sqrt{m}K}^{\sqrt{m}K}\gamma\omega^{2}\overset{.}{z}_{GR}^{r}%
\cos\left(  \omega t\right)  dt. \tag{C4}%
\end{align}
After substituting $\overset{.}{z}_{GR}^{r}(t;t_{0},m)$ from Eq. (16) into Eq.
(C4), using the Fourier series of the elliptic function $\operatorname*{dn}%
\left(  \cdot;m\right)  $ [18], and using standard tables of integrals [28],
one obtains the average energy over the period $T^{r}$ as a function of the
initial phase difference $\left(  \omega t_{0}\right)  $ between the harmonic
excitation and the GR solution%
\begin{align}
\left\langle \frac{dE}{dt}\right\rangle _{T^{r}}\left(  t_{0}\right)   &
=8\eta\sqrt{m}E^{\ast}\mp\frac{2\pi\eta\gamma\omega\sin\left(  \omega
t_{0}\right)  }{\cosh\left[  \omega\sqrt{m}K^{\prime}\right]  }\nonumber\\
&  \mp\frac{2\pi\gamma\omega^{2}\cos\left(  \omega t_{0}\right)  }%
{\cosh\left[  \omega\sqrt{m}K^{\prime}\right]  }, \tag{C5}%
\end{align}
where the upper (lower) sign in Eq. (C5) refers to counterclockwise
(clockwise) rotations. From Eq. (C5) one sees that a \textit{necessary}
condition for $\left\langle \frac{dE}{dt}\right\rangle _{T^{r}}\left(
t_{0}\right)  $ to change sign at some $t_{0}$ is written
\begin{equation}
8\eta\sqrt{m}E^{\ast}\leqslant\sqrt{\left(  \frac{2\pi\gamma\omega\eta}%
{\cosh\left[  \omega\sqrt{m}K^{\prime}\right]  }\right)  ^{2}+\left(
\frac{2\pi\gamma\omega^{2}}{\cosh\left[  \omega\sqrt{m}K^{\prime}\right]
}\right)  ^{2}}, \tag{C6}%
\end{equation}
where the equals sign in Eq. (C6) yields the threshold amplitude $\gamma
_{th}^{r}$ given by Eq. (31).

\end{document}